# Wise Computing

## Towards Endowing System Development with True Wisdom

(Preliminary Version; January 2015)

David Harel, Guy Katz, Rami Marelly and Assaf Marron

The Weizmann Institute of Science, Rehovot, Israel

**Abstract.** Encouraged by significant advances in algorithms and tools for verification and analysis, high level modeling and programming techniques, natural language programming, etc., we feel it is time for a major change in the way complex software and systems are developed. We present a vision that will shift the power balance between human engineers and the development and runtime environments. The idea is to endow the computer with human-like wisdom – not general wisdom, and not AI in the standard sense of the term – but wisdom geared towards classical system-building, which will be manifested, throughout development, in creativity and proactivity, and deep insights into the system's own structure and behavior, its overarching goals and rationale. Ideally, the computer will join the development team as an equal partner – knowledgeable, concerned, and responsibly active. We present a running demo of our initial efforts on the topic, illustrating on a small example what we feel is the feasibility of the ideas.

## 1      What is the problem?

Major advances in languages, tools and methodologies have improved our ability to develop reactive systems, but the task remains very difficult, expensive and error prone. Deliverables can fail, bringing in their wake disastrous results that are far worse than merely exceeding budgets and time schedules. One of the key reasons for this gap is that the sheer complexity of many kinds of reactive systems keeps growing, and increasingly prevents the human mind from managing a comprehensive picture of all relevant elements and behaviors of the system and its environment. Also, of course, in general, the state-explosion problem prevents us from exhaustively analyzing all possible behaviors of the system.

While each new generation of languages and tools helps tackle the above issues, the solutions reach their limits, and resolving the great difficulties of developing reliable reactive systems remains a major, and critical, moving target.

## 2      Oh, wouldn't it be nice if . . .

Our vision is to bring about a major change in the way complex software and systems are developed, by shifting the power balance between the human engineers and the development and runtime environments**.** Encouraged by the many recent advances in several fields, we propose a novel computing paradigm, whereby we turn the development environment, as well as the final system itself, into a much smarter, proactive, creative and interactive stakeholder in the development process and during execution and maintenance. Ideally, the computer will join the development team as an equal partner – knowledgeable, concerned, and active. Implementing such a vision would require one to harness many ideas, algorithms, techniques and tools, developed by researchers over the years.

As a small example, a wise environment will be able – on its own – to identify a missing requirement of an alarm for the loss of communication with a temperature sensor, explain that without such sensor data the system will keep a certain door shut to reduce risk of fire, and recommend that before opening the door manually, one should check for fire on the other side. And all this communication will be carried out in very high level fashion.

As another example, consider a control system being developed for a chemical plant. Suppose we could ask the system "what happens if the temperature sensor malfunctions during this-or-that procedure?", using terminology that may or may not exist at the time within the system. And suppose that a wise development environment could not only answer such questions, but proceed to check system behaviors and properties, including such that were not explicitly specified by humans. It would then alert us proactively, for example, to the fact that some new functionality might affect a previously specified critical behavior; e.g., that an alarm device previously dedicated to one situation is used also for another situation, possibly confusing human operators.

Moreover, suppose that years after initial development, we wanted to add functionality to deal with a new kind of remote safety monitoring standard. Wouldn't it be wonderful if the system could be told about this in an easy-to-use visual fashion, or in natural language and it would indicate to us (in a similar language) the affected parts in the system? Indeed, even the mere confirmation that such functionality does not yet exist would already be extremely valuable.

We have felt for some time now that, given the great amount of relevant existing research, the time seems ripe to seriously address the aspirations voiced above. And indeed, for several months we have been working hard on the broad concept underlying them, which we term *wise computing*. Interestingly, just as we were getting ready to finalize this initial paper on the topic, past ACM president Vinton Cerf published a *CACM* column (Jan. 2015) expressing the hope that people might consider working on (parts of) this kind of vision.

## 3 A lot has already been done

There is a tremendous amount of past research that is key to addressing the ideas voiced in this paper. This includes work on new programming paradigms and languages, as well as model-driven engineering methods and tools, and domain-specific languages and architecture frameworks. Specific technical areas include requirements engineering, test generation, verification, synthesis, specification mining, code analysis and refactoring, model queries, ontologies, knowledge representation and engi-neering, learning, natural language processing, human machine interface, planning, multi-agent techniques, search and recommender systems for software engineering, and more.

Being a vision paper, we shall not provide here a detailed review of these fields and of their main relevance to our ideas. In lieu of this, we are preparing a supplementary document on related work, which will be continuously updated and enriched, and which will be placed here:

***http://www.wisdom.weizmann.ac.il/~harel/wisecomputing***.

The work our group has done on an earlier and more modest dream, termed at the time *liberating programming*, is in a way a precursor to the present vision. In a 2008 paper, we suggested that, using a scenario-based development approach, programmers could be brought a lot closer to the way we "program" others; i.e., the way we explain things to our children, employees, students, etc., in order for them to attempt to understand what we had in mind and act accordingly. We have termed a generalized version of this approach *behavioral programming* (BP).

## 4 So what is missing?

Well, even with all the work done over the years, the computer being programmed is still that passive obedient servant, doing what it is told. Yes, the system and its development environment have been made smarter, understanding programming on higher levels of abstraction, and being able to rely on examples and adapt to changes in its environment. Yes, a system can now deal with more natural, multi-modal, incremental specifications of inter-object scenarios, in a style that is closer to the way humans think about reactive behavior. And yes, we are better at analyzing, testing, verifying, and perhaps even synthesizing reactive systems. But the process is still one-way: it is still *us* instructing *it*, the computer, while all *it* does is make an effort to understand what we had in mind (assuming that we ourselves have a clear picture thereof...).

These capabilities are still very different from how humans interact when they plan new systems, discuss how to deal with new tasks, or teach or coach others. Such activities are two-way, and involve far more elaborate relationships and complex and subtle processes that take into account deep knowledge, both tacit and explicit, and many, often conflicting, desires, needs and constraints. All sides to the collaboration contribute their knowledge and wisdom to the effort, in line with the dynamics of the project's development, and very often at their own initiative, unprompted.

And so, we believe that a major breakthrough is called for, fundamentally changing the relationship between the human engineer and the development tool/environment. It is about time we considered treating the computer we are programming as an entity capable of far deeper wisdom about behavior and development than is currently thought. Such an entity would interact with us wisely, like a colleague. It could respond to our needs with knowledge and, utilizing extensive computing power it would have "under the hood", could proactively help in the variety of tasks that constitute the development process of the actual desired system.

## 5    Here is what we would like

The *wise development suite*, or the WDS, as we shall call it, should become a creative and proactive stakeholder, perhaps even a leader, in the development process (see Fig. 1). It will possess aspects of human-like wisdom – not general wisdom, but wisdom geared towards system-building. This will be manifested in it initiating discourse and actions based on deep insights into the system's own structure and behavior, and its overarching goals and rationale.

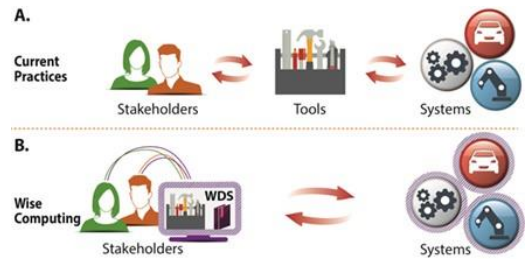

**Fig. 1.** The WDS (wise development suite) joins the stakeholders and developers as an equal, knowledgeable, proactive partner throughout the lifecycle of system and software engineering.

The WDS should use relevant knowledge (both general and domain-specific), which may not be captured explicitly within the system itself, and it should have the ability to distinguish between what is essential and what is of only secondary importance. It will participate in the elicitation, formalization, validation and iterative enrichment of requirements, thus helping to increase confidence in the semantics and intention of the requirements, and establishing their consistency and inter- relationships. And clearly, in the spirit of almost 30 years of model-driven development, the WDS will also be central to the ability to directly execute/simulate those requirements and/or translate them into running application code.

We would like all of this done so that the constant "worrying" about whether things will work right will flow in both directions between the human and the system.

The interaction part of the WDS will be characterized by an ability to use various levels of abstraction, suitable for the topic and for the human and machine participants. It will use visual representations, examples, pseudo- and conventional code, etc., enabling understanding by a wide range of stakeholders.

A key capability will be the use of natural language in both directions. Indeed, despite much work on controlled natural languages for requirements and program specification, we are still far from the point where we can automatically read and parse requirements specified in a way that is natural and accessible to humans, and from them create a correct formal specification.

The WDS will be able to explore functionality and behavior both exhaustively and under various "what-if" conditions and use-cases, communicating on multiple levels. Throughout development and maintenance, the computer will thus be constantly investigating itself, in a sort of self-aware fashion. On the one hand, it will enhance models and documentation to help human understanding, and, on the other hand, it will be constantly "concerned", like a worrisome colleague. It will detect problems in the system under development, including bad behaviors, conflicting behaviors, goals and requirements that are not met, missing specifications where human instructions are needed, inefficiencies in execution, and unneeded complexities in specification and implementation. The WDS will then initiate and propose changes and enhancements, including, if needed, refactoring and restructuring to help resolve issues.

Returning to the chemical plant example, the WDS will be able – on its own – to detect a missing alarm or safety check, detect that another alarm has no time limit or reset facility, identify a potential undesired valve opening, predict what will happen (and why) following changes proposed by programmers or by the WDS itself, and identify ways to simplify the decision of when to open the valve.

At runtime, the system will be able to interact with users and with other systems in order to explain past behavior and allow the user to influence future behaviors, creating a new dimension of control. For example, a door in a chemical plant or an airplane will be able to explain to a human why it is presently closed, what will happen if it is manually opened, and discuss in detail sensor information and alternative sequences of manual and automated actions associated with opening and closing it.

The system's knowledge will come from broad and multi-faceted sources. It will not be limited to the project at hand. Rather, it will reflect cumulative organization-wide and industry-wide knowledge about similar systems, about the problem domain with its standards and practices, and about system engineering in general. Such knowledge will keep evolving and be constantly disseminated and expanded.

Our goal is not full automation that creates a black-box system from natural specifications. Rather, we envision the creation of an environment that serves as a proactive stakeholder, who can share insightful information, criticism and suggestions with humans throughout the system's life span. This can benefit even systems initially created by full automation.

The most immediate benefit of a wise computing suite will be, of course, a significant reduction in the development time and cost of complex systems, and will result in improved system quality. Run-time wisdom, combined with ubiquitous run-time human interaction, will increase user and regulator confidence in systems, further expanding development and adoption. And over and above all of this, we believe that in the farther future we will experience new dimensions of innovation, as rich new capabilities and new ranges of safety will be initiated (nay, invented!) by wise systems, rather than only by humans.

How will it do all this? Well, some of our ideas on this are explained in the next section and some capabilities along these lines (albeit a very modest collection thereof) are present already in our initial work, demonstrated in Section 7.

## 6      Towards getting it done

Our dream is not about artificial intelligence in the standard sense of the term. It is also not about programming or verification *per-se*, or about automating system development. It is about turning today's one-way engineering process of developing complex systems into a wise, interactive two-way endeavor.

The aforementioned work in our group on liberating programming, and the BP paradigm, enables humans to build a system in a way that is aligned with how they perceive its desired behavior and functionality. While these developments can serve as supporting building blocks in wise computing, they are far from enough. First, the feasibility of wise computing does not prescribe any particular approach to programming (though our initial effort described in Section 7 was carried out within the BP framework). Second, we propose that the focus shift to what the system itself and its supporting environment, the WDS, know about the system and how they can use this knowledge.

To accomplish this, we believe that a coordinated and interwoven 3-prong research effort is needed, building upon the most powerful existing means: (i) new ways for representing and structuring knowledge about systems, (ii) new algorithms for analyzing and understanding systems, and (iii) new approaches to human-computer discourse about systems.

These three, even if carried out to perfection, will not be adequate on their own. A major part of the work on wise computing will have to deal with putting together a comprehensive system development tool, with a corresponding methodology, which will include all three facets of the approach, and which the engineers should find extremely productive and reliable, as well as pleasant and intuitive to use. Moreover, such a wise computing tool would have to be relevant throughout the lifetime of even very large and complex systems. Hence, scale-up will become a crucial issue in all parts of the effort, as will a sort of legacy-retaining nature.

### 6.1 Representation and structuring

Wise computing will require us to develop powerful ways, accompanied by rigorous semantics, to represent behavior, structure, and other related system information in models and in executable specifications, adding to and complementing those presently in use in research and in mainstream development practices. Such means should include the ability to express multiple coexisting levels of abstraction, and interrelationships between and among components and behaviors, as well as ways to define preferences and priorities in application-agnostic ways.

This would entail having adequate methods to represent the integrated management of different types of system properties, such as behavioral and structural, discrete and continuous, deterministic and stochastic, object-centric and inter-object. The interrelationships between behaviors would have to include creating cross-behavior awareness (e.g., awareness of state or properties thereof) without adversely affecting encapsulation, as well as to accommodate the dependencies created in natural processes when humans describe behavior incrementally. A common example is the case of rules and exceptions, such as "whenever the system does X, it should also do Y, and not do Z unless condition C holds".

Such formalisms could extend concepts from approaches such as scenario-based/behavioral programming, as well as from statecharts, aspect orientation, contract-based design, behavior-intention-priority (BIP), feature oriented software development, and the subsumption architecture, in order to combine natural discourse with robust design. The abstractions will allow natural, yet precise and verifiable, refinement. And it is here that research on natural language processing in the context of programming comes in too.

Knowledge bases and ontologies containing general software engineering and domain-specific and project-specific knowledge, as well as techniques to access and utilize them, will have to be significantly enhanced, so that they will serve as support for the WDS. The required information will be built from existing ontologies, industry standards and reference documents as well as information provided by project stakeholders, and/or fed through crowd sourcing and search-and-recommender systems.

An important aspect of the approach is to allow both gaps and overlaps in the specifications or executables. Overlaps play a key role in accommodating multiple stakeholders and gaps are a natural product of a system's evolution. Using its structuring capabilities, the WDS will be able to identify gaps, either helping to eliminate or reduce them, or facilitating "living with them" for as long as possible.

### 6.2     Analysis and comprehension

Suitable ways will have to be devised to render the WDS capable of "understanding" what a system does or needs to do, and to explain this to the stakeholders. Such algorithms will undoubtedly use (and modify or extend as needed) state-of-the-art techniques in many of the fields mentioned earlier, like verification, synthesis, and learning. In particular, there will be algorithms that deal with exploring predicted behavior, identifying bad behaviors, conflicts and underspecification, and proposing necessary corrections.

The design will be tied to the requirements and to the general and domain-specific knowledge, which will enable insightful discourse regarding requirement tracking and impact analysis, as well as detection of missing requirements or design omissions.

One of our goals is to be able to identify complex behavioral dependencies and causalities, which can be simplified for clarity or otherwise modified for better performance. The WDS will then drive such changes by streamlining individual components and by identifying and consolidating otherwise hidden emergent patterns. Ideally, the patterns will not be just syntactical, but will emerge from understanding the relationships between requirements, functions and structures, representing concepts that are meaningful to human stakeholders. The focus will be on offering design simplification in view of both the individual changes and the composite system.

An obvious problem with almost any kind of analysis is computational complexity and the state explosion problem. We will make an effort to enhance the ability to deal with this by working on multiple levels of abstraction and exploiting modularity. As to the latter, we will take advantage of the modularity offered by the programming medium involved (in our case, BP) to yield, when possible, more efficient compositional verification. We already have some preliminary results (as of yet unpublished) in applying SMT solvers to behavioral programs, which support this.

### 6.3     Human-computer discourse

The interaction with the WDS must be powerful, yet readily usable by a variety of stakeholders. There appears to be a need for a new high-level interaction language, utilizing, where applicable, ideas from previous work on visual formalisms and GUI-based play-in, as well as on natural language, temporal logics, and the variety of existing collaboration languages.

Both the computer and the user will be able to refer to system-generated and user-generated insights and recommendations. These include goals and requirements, reasons for achieving or not achieving them, various system properties, progression and results of event sequences.

Interactions will be two-way, initiated by the human or the WDS, and, as discussed earlier, it will be possible to carry them out in natural language too. They will also be hierarchical, so that interaction elements will be composable into higher level ones, and they will be relevant in all stages of the project.

Since the WDS is expected to deal also with the system as it evolves through iterative development and maintenance, it will have to maintain a great deal of past information, such as answers to queries and properties previously verified or documented. Thus, what is sometimes referred to as the human's mental model of the system under development will be enhanced in the wise computing world by this grand collection of evolving maintained artifacts, common to the human engineers and the WDS.

## 7 Demonstrating our initial work

We end this initial paper on wise computing by illustrating the feasibility of some of its ideas. A pre-recorded demo of our preliminary and very modest wise computing system, over-voiced with explanations, can be downloaded from here:
***http://www.wisdom.weizmann.ac.il/~harel/wisecomputing***.

We emphasize "preliminary" and "modest" for several reasons. First, despite prior work in many fields, we believe that much of the vision described above is not yet doable in its desired generality. For example, the demo does not yet illustrate what-if queries or natural language input, nor does it work on multiple levels of detail. In addition, it involves an extremely simple example, and the numerous issues of scale-up are yet to be dealt with.

Still, the work we have already done in constructing the "mini-WDS" that underlies the demo was far from trivial, and it exhibits new ideas beyond what we have seen in current system development tools. In any case, it seems to point to the feasibility of wise computing.

As explained orally in the demo itself, the tool involves three new components, over and above the programming language itself (BPC; i.e., behavioral programming in C++, allowing the waiting, requesting and blocking of events.[1]) and the other existing tools we use, such as a model checker and an SMT solver. We call these components *the three sisters*: Athena, Regina and Livia.

Very briefly, Athena, the wise one, works proactively, in an offline fashion, during development. On her own, she uses formal tools such as a model checker to analyze things she feels to be relevant, and produces mathematically accurate conclusions that are valid for all runs.

---

[1] D. Harel and G. Katz, "Scaling-Up Behavioral Programming: Steps from Basic Principles to Application Architectures", *Proc. 4th SPLASH AGERE! Workshop*, 2014

Regina, more regal than her sisters, and also working offline, actually runs the system proactively many times, collecting statistical information as she goes. In what can be seen as a form of specification mining, she then attempts to reach interesting conclusions, and makes suggestions to the user. Her conclusions may not be valid in all runs, but they have the advantage of reflecting numerous executions, and thus can capture what will happen in typical runs.

Finally, Livia works in a live online fashion, monitoring the system as it runs and helping the user or designer handle all kinds of situations in real time. She also supports integrating executed scenarios into the system as test-cases (i.e., playing them in), applying coverage criteria to check for test redundancy.

The three sisters can call upon each other for help – e.g., Livia can call Athena to verify whether an observed property indeed always holds.

In order to illustrate the inner workings of a future wise computing system, here is a high-level description of parts of Athena's current algorithm. More details, as well as similar descriptions of Regina and Livia, will be published separately.

*Identifying threads*: Athena parses the program's source files, seeking the threads.

*Constructing state graphs*: Athena runs each thread in a sandbox to construct a representation of its states and transitions.

*Partitioning*: A thread "cares about" an event if in many of its states it affects or is affected by it. This definition induces a partitioning of the threads into modules, each including those that care about the same event. Through thread partitioning, Athena can recognize parts of the code that deal with the same facet of the system.

*Pattern matching*: Here Athena identifies patterns in the threads' state graphs and extracts relevant information. Supported patterns currently include cyclic threads, array components, periodic programs, sensor/actuator threads, and more.

*Deciding what to check*: Some properties are general (e.g., deadlock freedom) while others are program-specific. If Athena detects an array component in the program, she might decide to check that information stored in the array eventually gets read. In a periodic program she can check that no task deadlines are breached.

*Checking and recommending*: Built atop the BPC framework, Athena (and also Regina and Livia) has interfaces to several verification tools that we have developed in the last few years within the BP and BPC framework: a special model checker, an abstraction-refinement mechanism, automatic correction tools, synchronization relaxation tools, and an SMT solver. Based on the property at hand, these tools can be invoked. Athena presents to the programmer the conclusions she reaches, and will attempt to suggest ways to correct problems she found (often by synthesizing or modifying program threads).

## 8   Conclusion

We are excited by the wise computing vision outlined here. Clearly there is much previous work that is deeply relevant to it (see the supplementary document mentioned earlier), and which will have to be adapted and extended before it can be incorporated into a serious and useful wise computing development suite. Still, we hope that the paper and accompanying demo will inspire other research groups to become part of an effort to bring wise computing to fruition. We plan to carry out our own work on the topic using an open-source and extendible conceptual architecture and infrastructure.

## Acknowledgements

This work was supported by a grant from the Israel Science Foundation and by the Philip M. Klutznick Fund at the Weizmann Institute. We thank Orna Kupferman, Moshe Vardi and Gera Weiss for their valuable comments on the ideas expressed here.